# All you need is data: the added value of National Access Points as backbone European ITS data exchange infrastructures

**Chrysostomos Mylonas[1], Maria Stavara[1], Evangelos Mitsakis[1]**

[1]Centre for Research and Technology Hellas – Hellenic Institute of Transport
E-mail: chmylonas@certh.gr, mstavara@certh.gr, emit@certh.gr

**Περίληψη**

Τα Ευφυή Συστήματα Μεταφορών (ΕΣΜ) είναι ζωτικής σημασίας για τον ψηφιακό μετασχηματισμό των μεταφορών. Η Ευρωπαϊκή Επιτροπή έχει επιβάλει την ίδρυση Εθνικών Σημείων Πρόσβασης (ΕΣΠ) σε κάθε κράτος μέλος, τα οποία λειτουργούν ως κοινές εθνικές διεπαφές για την ανταλλαγή δεδομένων ΕΣΜ. Παρόλο που έχει σημειωθεί πρόοδος στην προτυποποίηση των δεδομένων των ΕΣΠ, η ενσωμάτωσή τους με τις λειτουργικές πρακτικές ΕΣΜ παραμένει περιορισμένη. Η παρούσα εργασία παρουσιάζει πέντε περιπτώσεις μελέτης που θα χρησιμοποιούν δεδομένα από το ΕΣΠ. Η πρώτη μελέτη περίπτωσης περιγράφει ένα εθνικό εικονικό κέντρο διαχείρισης κυκλοφορίας που προσφέρει σχεδόν πραγματικού χρόνου οπτικοποιημένους δείκτες για την υποστήριξη λειτουργιών κυκλοφορίας στους αυτοκινητόδρομους. Η δεύτερη μελέτη αφορά την παροχή μηνυμάτων μέσω συνεργατικών ΕΣΜ σε χρήστες οδού. Η επόμενη μελέτη επικεντρώνεται σε μια πανευρωπαϊκή διεπαφή που παρέχει οπτικοποιημένα τα διαθέσιμα δεδομένα των ΕΣΠ. Τέταρτη μελέτη περίπτωσης αποτελεί η ψηφιοποίηση σχεδίων διαχείρισης κυκλοφορίας μεταξύ των αντίστοιχων κέντρων διαχείρισης. Τέλος, η πέμπτη περίπτωση οραματίζεται μια τεχνική διεπαφή που συνδυάζει τα δεδομένα κυκλοφορίας ΕΣΠ με μετεωρολογικές πληροφορίες και παρέχει σχετικούς δείκτες με τις επιπτώσεις ακραίων καιρικών φαινομένων στο δίκτυο.

**Abstract**

Intelligent Transport Systems (ITS) are crucial in the digital transformation of transportation. The EC mandates the establishment of National Access Points (NAPs) in each Member State, serving as common national interfaces for ITS data exchange. While progress has been made in standardizing NAP data, integration with operational ITS practices remain limited. This paper presents five NAP use cases from the NAPCORE (National Access Point Coordination Organization for Europe) CEF program. The first one outlines a National Virtual Traffic Management Center offering real-time visualized KPIs supporting motorway traffic operations. The second focuses on NAP-enabled Cooperative ITS and dynamic traffic management services. Next use case involves a Pan-European interface, providing visualizations of data availability. The fourth use case enhances the digitization of traffic management plans, among different TMC. Finally, the fifth use case demonstrates a technical interface combining NAP traffic data with meteorological information for KPIs on extreme weather impacts on traffic.

## 1. Introduction

The European Commission (EU) has proposed the deployment of National Access Points (NAPs) for the exchange and distribution of ITS data. This proposal was made within the context of the Delegated Regulations No. 885/2013, 886/2013, 962/2015, and 1926/2017 that supplement the ITS Directive (2010/40/EU). In the context of that Directive, a NAP may be viewed as a single digital platform at a national level, which provides access in a centralized or decentralized manner to properly formatted ITS-related data accompanied by appropriate metadata. In an effort to strengthen the NAPs, the European Union has funded several projects and standardization mechanisms. These endeavors are primarily directed towards achieving data harmonization and alignment with specific standards and formats. Nevertheless, NAPs have not been fully leveraged, as their data does not seem to be directly connected to ITS services and applications. The current paper aims to shed light on the significance of NAPs as fundamental European ITS data exchange infrastructures, emphasizing on their added value use cases enabling their practical exploitation and further harmonization across different Member States. With the growing complexity of urban areas and the increasing demand for efficient mobility solutions, ITS services have emerged as a key enabler for smarter, safer, and more sustainable societies. Hence, it is important to recognize the benefits arising from these services given their increased penetration rate and their implications on the quality of life.

## *2. Empowering Mobility: The Significance of a National Access Point*

Before delving into the significance of a NAP, it is necessary to provide a clear understanding of the real concept behind these data platforms and their contribution to the entire data chain. In essence and as already mentioned in the introductory part, a NAP serves as a central point where ITS-related data are gathered and made available in the form of datasets. The data resources included in each dataset shall be structured and comply with specific rules and standardized formats. These formats are to a great extent determined by their content or their relevance with the data categories described within the Delegated Regulations supplementing the ITS Directive. For instance, real time traffic-related information should comply to the standards developed by the DATEX II community, while static multimodal information should comply to the NeTEx standard developed by CEN. These data resources may be hosted on NAPs using a centralized approach, or alternatively, the NAPs may provide links for accessing data resources that are hosted on external databases and systems (decentralized approach). The former approach bears more similarity to a database concept, while the latter approach resembles a metadata repository concept. In any case, NAPs provide a rich list of benefits that are useful to explore. Characteristically, through the provision of metadata that conforms to specific terminology, data consumers are able to readily grasp the meaning, the structure, the nature and other interrelationships among the rest of the accommodated data records. It should be noted that metadata serves a dual purpose of enhancing data comprehensibility for both humans and machines. Nonetheless, this is only feasible if metadata conforms to the required standards and can be properly formatted using suitable profiles and control vocabularies, such as DCAT-AP. Furthermore, by consolidating data into a common access point (centralized or decentralized), the discoverability of data by human users is significantly enhanced. This becomes particularly valuable when data resources are scattered across multiple portals at a national level. In addition, when data is spread out in separate platforms, it becomes inherently challenging to have confidence in its quality. On the contrary, NAPs counteract this challenge by gathering all data in one place and offering consistent information regarding its quality (including KPIs). Hence, any uncertainties that may arise are alleviated. Finally, NAPs improve the interoperability of data. As previously stated, by implementing specific protocols and accommodating Application Programming Interfaces (APIs), NAPs establish a robust framework that facilitates automated data exchange, including dynamic data, and encourages machine-to-machine communication. However, even if NAPs provide access to a rich set of information and data, this does not necessarily translate to the creation of added value for the end users of or other actors involved in the operations of transport systems. "Added value" is herein understood as the utilization of NAP data and information with the provision of valuable services. Therefore, without the integration of ITS services, the data hosted on NAPs remains static and lacks practical significance.

## *3. Unlocking the future: Exploring five use cases in traffic operations*

This chapter focuses on exploring some use cases in the transportation field. The sub-sections will dive into five distinct use cases, namely the national virtual traffic management center, NAP enabled C-ITS services, pan-European data visualization platform, exchange of traffic management plans, and resilience observatory. By examining these use cases, this chapter aims to provide insights into how NAP data can be used to advance the efficiency, safety, and resilience of transportation systems as well as of the associated operations management processes.

*3.1 National Virtual Traffic Management Center*

The concept of virtual TMCs (vTMC) is not new. During 2014, the US FHWA published guidelines on how to develop a vTMC. In this document, a vTMC is addressed as a hub of several freeway management systems, wherein data about freeway traffic operations are concentrated, fused with other data sources, and processed/synthesized to extract information to be distributed to several stakeholders, including transport agencies (Lukasik et al., 2014). Among others, the staff of physical TMCs utilize the vTMC and its provided information to monitor (freeway) traffic operations and initiate or even terminate traffic control strategies called to affect/ameliorate traffic flow conditions. In a similar manner, more than one operator or agency can coordinate their responses in anticipation of specific circumstances and incidents. According to the US FHWA published guidelines, there are specific steps until the implementation of a vTMC, providing a structured approach to ensure the efficient operation of a vTMC. The current use case involves the development of an interactive platform (dashboard) supporting the operational monitoring of traffic conditions prevailing along the Greek national motorway network. The dashboard will be fed with data from the Greek NAP and will provide relevant KPIs in a numerical and visualized manner. These KPIs will provide valuable insights into traffic flow, congestion levels, average speeds, the impacts of on-going incidents and events, and other important aspects. The overall concept of the vTMC and its integration with the NAP revolves around improving traffic management efficiency, enhancing safety, and providing a better experience for road users. By leveraging real-time data from the NAP, the vTMC facilitates proactive decision-making and allows for timely interventions to address traffic congestion, incidents, or any potential disruptions. The dashboard's visualized representation of KPIs enables operators and stakeholders to quickly grasp the state of the motorway network and take appropriate actions. It is important to mention that all motorways can upload their datasets, ensuring that there is a common and up-to-date picture of the entire road network. This functionality enhances coordination and collaboration among different motorways, facilitating effective management of traffic conditions. The platform's visualization capabilities offer a comprehensive overview of the prevailing traffic conditions, allowing operators to monitor the network and identify potential bottlenecks or congestion areas. As can be perceived, the current use case holds a significant added value since many actors are benefit, including motorway operators, road users, and road authorities. On the one hand, motorway operators gain access to timely and accurate information about traffic conditions, allowing them to make decisions regarding traffic management strategies, incident response, and resource allocation. The visualization capabilities of the dashboard enhance situational awareness, enabling operators to quickly identify areas requiring attention and take proactive measures. On the other hand, other actors, such as emergency service providers, can receive real-time updates on traffic conditions to optimize response times and navigate efficiently through the network. Hence, in an overall perspective, the vTMC scenario improves operational efficiency, enhances safety, and delivers a better travel experience for road users across the Greek national motorway network. Figure 1 presents the overall concept and high-level architecture of the vTMC scenario.

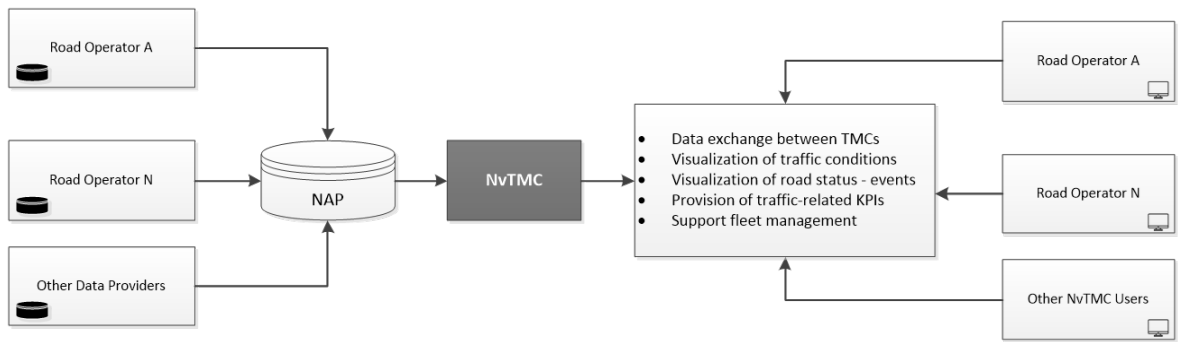

*Figure 1:* High-level architecture of national virtual traffic management center

### 3.2 NAP Enabled C-ITS Services

The next use case revolves around the provision of Cooperative Intelligent Transport Systems (C-ITS) messages to road users, enabled by data published on the Greek NAP. Specifically, this implementation focuses on delivering In-Vehicle Information (IVI) messages to vehicles based on the content broadcasted by road operators through Variable Message Signs (VMS). In this use case road operators utilize the NAP as a central data repository to publish relevant information for the road network related to traffic conditions, road hazards, diversions, weather conditions, and other pertinent information. The data from the NAP serves as a source for generating IVI messages by C-ITS service providers. The IVI messages, derived from the data published on the NAP and broadcasted through VMS, are then communicated to vehicles equipped with C-ITS capabilities. These messages are received by the vehicles' onboard systems and presented to the drivers through human machine interfaces, providing them with real-time and contextual information about the road conditions ahead. In this respect, road users gain access to up-to-date and relevant information while driving, allowing them to make decisions and adapt their route or driving behavior accordingly. They will also receive timely alerts about traffic congestion, accidents, road closures, or any other road-related information that may affect their journey. This leads to improved travel efficiency, reduced travel time, and enhanced road safety. This use case is expected to be highly useful as it enhances communication between road operators and road users. The overall concept and high-level architecture of that use case is depicted in Figure 2.

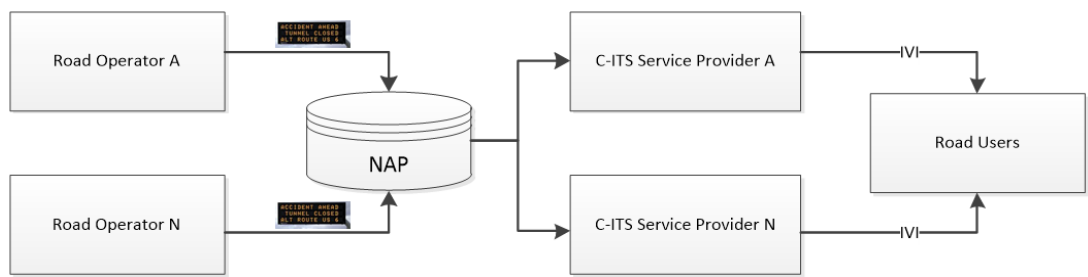

*Figure 2:* Normal flow of NAP enabled C-ITS services

### 3.3 Pan-European data visualization platform

Even though the concept of a Pan-European data visualization platform is not new, the development of such platform leveraging data from several European NAPs, is yet to come. With the advancement of standardization activities and the increasing availability of open data initiatives in many European countries, it is becoming more feasible to gather, standardize, and integrate datasets from various sources into a centralized platform. This platform aims to host and visualize datasets containing commonly formatted information from multiple European countries. For instance, a concrete example could be the visualization of datasets including the speed of the TEN-T parts of the European motorway network. In that case, the platform would provide valuable insights into the road conditions across different countries and facilitate cross-border decision-making for various stakeholders. By visualizing road speeds in many countries, transportation authorities can identify congestion-prone areas, analyze traffic patterns, and plan infrastructure developments accordingly. This can lead to optimized traffic management, reducing travel times and improving road safety as well. The current web-based platform will benefit road operators and road authorities from several European countries by obtaining a simultaneous picture of the on-going traffic conditions along the TEN-T motorway network. Figure 2 demonstrates the concept and high-level architecture of pan-European data visualization platform use case.

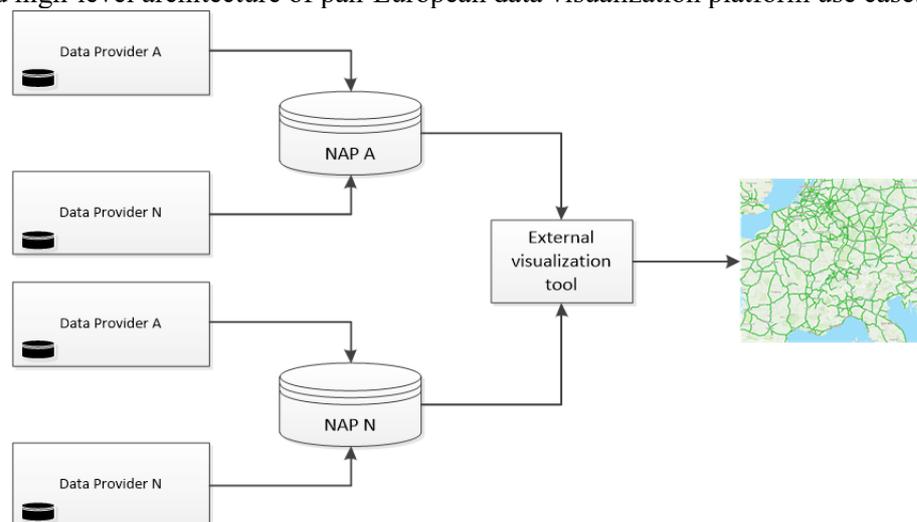

*Figure 3:* Concept of pan-European data visualization platform

## 3.4 Exchange of traffic management plans

Traffic management plans, as described in the relevant literature, are comprehensive strategies and procedures designed to effectively manage and control traffic flow in various situations. These plans typically involve the implementation of temporary measures, including lane closures, diversions, and speed restrictions, to ensure the safety in work zones, construction areas, special events, or emergency situations (Kurzhanskiy et Varaiya, 2015). Traffic management plans have as an ultimate goal to minimize disruptions, maintain traffic flow, provide clear communication to road users, and optimize the efficiency of traffic operations. They often involve coordination among multiple stakeholders, including transportation authorities and emergency services, to ensure a systematic and well-coordinated approach to managing traffic in a given condition and location. The current use case envisages the exchange of fully digitized traffic management plans among different Member States in Europe through NAP, thus enabling several stakeholders to submit, access, and update their traffic management intentions and measures. This facilitates seamless communication, collaboration and coordinated decision making between the aforementioned stakeholders. At a second phase, this use case can promote standardization and consistency by enforcing a uniform data format and protocols, facilitating that way the easier interpretation, analysis, and publication of traffic management plans. Moreover, the development of such a system will enable the storage and the analysis of historical data, which plays a crucial role in improving future planning and decision-making. By maintaining a repository of past traffic management plans, stakeholders can access valuable information about previous strategies identifying similar patterns and trends. The involvement of many countries in the exchange of such plans will enable the efficient coordination of traffic operations (especially across borders) during large-scale events. Finally, that concept is to facilitate the real-time sharing of traffic management plans and relevant updates when emergency situations affect multiple countries. This enables a coordinated response, efficient evacuation routes, and effective traffic diversion strategies.

*3.5 NAP-enabled resilience observatory*

The "NAP-enabled resilience observatory" use case plays a crucial role in monitoring and assessing the impacts of extreme weather events and other significant incidents on traffic operations at urban road and highway network level. By leveraging data from the Greek NAP and/or open weather APIs, this observatory enables the correlation of traffic flow patterns with weather conditions. This correlation helps identify the specific effects of weather events on traffic congestion, road conditions, and overall transportation performance. The observatory serves as a valuable resource for gathering historical evidence and generating visualized KPIs related to the resilience of the transportation system. These KPIs provide insights into the relationship between weather events and traffic disruptions, enabling stakeholders to understand the magnitude and duration of the impacts. Collaboration among various stakeholders such as meteorological agencies or environmental monitoring stations is vital in the implementation of the resilience observatory. By fostering collaboration and information sharing among these stakeholders, the NAP-enabled resilience observatory enhances the understanding of the vulnerabilities and risks associated with extreme weather events and supports proactive decision-making for mitigating their impacts. This use case enables the development of targeted strategies and measures to improve the resilience of the transportation system, ensuring safer and more efficient travel even in adverse conditions.

## 4. Conclusions

ITS constitute a core driver for the future of the transport sector and especially for its adaptation to the new digital era. However, while ITS technologies offer immense potential, their impact is realized when they are seamlessly connected and utilized in real-world scenarios. It is not just about collecting and analyzing data; the true value of ITS lies in its application and integration into existing transport systems and practices. To drive the future of the sector, it is crucial to establish robust connections between ITS data and practical applications. This is exactly the mission that NAPs are called to accomplish. The five NAP use cases exemplify the diverse range of applications and benefits that NAPs can offer. In essence, the presented use cases highlight the importance of NAPs as critical infrastructure for data exchange and collaboration in the ITS domain. They showcase the potential of NAPs to enhance traffic management, support decision-making processes, improve coordination, and increase resilience in the face of various challenges. Continued efforts in standardization, integration, and utilization of NAPs are crucial for maximizing their benefits and advancing the digital transformation of transportation systems across Europe. Upon the time of the paper's presentation operational prototypes of the discussed use cases will be demonstrated.

## 5. References-Bibliography